# All-optical diode via nonreciprocal nonlinear absorption and interfacial charge transfer in two-dimensional van der Waals heterostructures

*Erkang Li, Jinhong Liu, Yanqing Ge, Mingjian Shi, Yijie Wang, Chunhui Lu\*, Yixuan Zhou\*, Xinlong Xu\**

Shaanxi Joint Lab of Graphene, State Key Laboratory of Photon-Technology in Western China Energy, International Collaborative Center on Photoelectric Technology and Nano Functional Materials, Institute of Photonics & Photon-Technology, School of physics, Northwest University, Xi'an 710069, China

E-mail: luchunhui@nwu.edu.cn (Chunhui Lu); yxzhou@nwu.edu.cn (Yixuan Zhou);xlxuphy@nwu.edu.cn (Xinlong Xu)

## Abstract

Nonreciprocity is fundamental to photonic and optoelectronic devices such as all-optical diodes for ultrafast optical signal processing. However, previous nonreciprocity is mainly based on linear optical response instead of nonlinear optical response based on recently developed two-dimensional (2D) van der Waals heterostructures. Herein, an all-optical diode prototype based on nonreciprocal nonlinear absorption and interfacial charge transfer is proposed and designed by both simulation and experiment based on ready van der Waals heterostructures. The giant saturable absorption from 2D MXenes (NbC) and reverse saturable absorption from 2D chalcogenides (GaS) play a synergistic role in the designed all-optical diodes, which is characterized by a femtosecond laser based Z-scan system. The comprehensive physical mechanism of this all-optical diode based on 2D van der Waals NbC/GaS heterostructure designed by simulations, is consistent with experiments under the consideration of both nonreciprocal nonlinear absorption and interfacial effect. This all-optical diode based on the 2D van der Waals heterostructure features the simplicity, scalability, stability, integration, and compatibility with the complementary planar fabrication technology, which can further extend and miniaturize the nonlinear photonic and optoelectric devices.

## TOC Graphic

In this paper, an all-optical diode prototype based on nonreciprocal nonlinear absorption and interfacial charge transfer is proposed and simulated. The 2D van der Waals NbC/GaS heterostructure is successfully applied for designing the all-optical diode in the experiment.

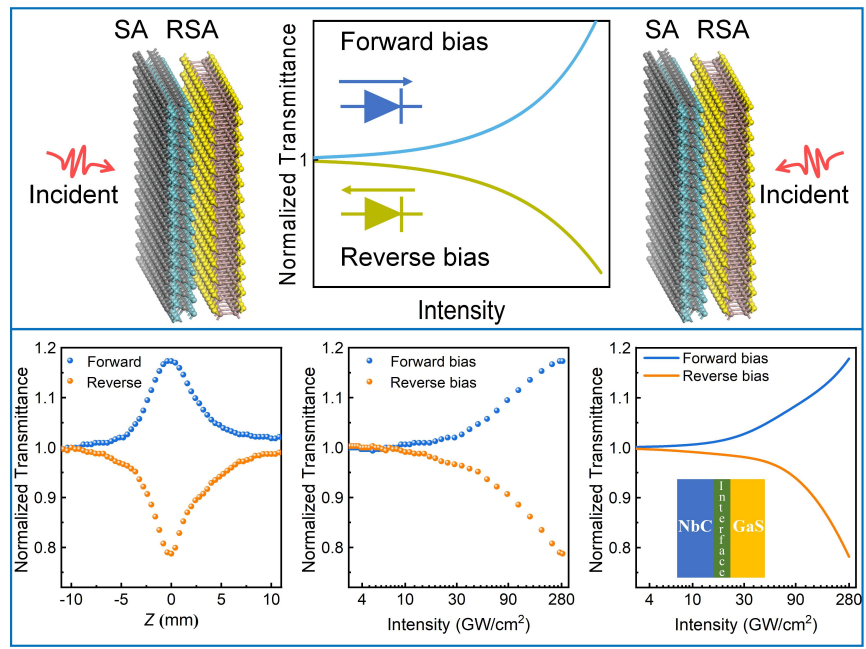

## 1. Introduction

Nonreciprocity, breaking the time-reverse symmetry, is fundamental to the operation in all-optical diode devices, which plays a significant and powerful role in ultrafast optical signal processing. However, previously reported diode responses were fabricated by means of opto-acoustic effect[1], magneto-optic effect[2], and all-silicon microrings[3] for nonreciprocity based on linear optical response, which require either complicated fabrication technology or external excitation sources such as magnetic field and acoustic field. The nonreciprocal nonlinear absorption, provided by the synergistic effect of both saturable absorption (SA) and reverse saturable absorption (RSA) materials, is an effective method to achieve all-optical diodes with simplicity and integration. Although early reported passive all-optical diodes mainly concentrate on the $C_{60}$[4-6] as a RSA material combining with other SA materials, recently developing two-dimensional (2D) materials is almost neglected. Besides, 2D materials including both MXenes and chalcogenides can usually exhibit large nonlinear absorption responses due to the 2D confinement effect[7-9], which is desirable to the nonreciprocity with 2D van der Waals heterostructures for all-optical diodes. Furthermore, there exists rich interfacial charge transfer[10, 11] in the 2D van der Waals heterostructures, which has a significant effect[12-14] on the performance of photonic devices. This calls for the comprehensive physical mechanism of all-optical diodes based on all-2D-materials different from previously reported diodes[4-6] with no account of the interfacial effect.

Although current 2D materials have exhibited excellent SA or RSA properties, hunting for large nonlinear absorption coefficients of SA and RSA materials is extremely necessary for new design and manufacture of all-optical diodes. The 2D metal materials generally exhibit broadband SA responses, which have been demonstrated in MXenes[15]. However, the function groups[16] in traditional MXenes introduce the uncontrollability of the SA properties in experiments. NbC as the simplest MXene with the high physicochemical stability[17] is desirable but less understood. Similarly, 2D chalcogenides such as $MoS_2$[18] and $WS_2$[19] usually can act as RSA materials due to the two-photo absorption. However, RSA material with wider bandgap, higher RSA response and wider wavelength response is an optimal choice. Compared with the transition metal dichalcogenides, GaS exhibits the wide bandgap (>2.2 eV)[20], which would be suitable for the ideal RSA material in the infrared region.

In this work, motivated by the strong SA of NbC and RSA of GaS, 2D van der Waals heterostructure based all-optical diode is designed and successfully fabricated. The NbC film and GaS film, prepared by liquid phase exfoliation and physical vapor deposition methods, are characterized by a femtosecond laser based Z-scan system, exhibiting large nonlinear absorption coefficients of -282 cm/GW and 812 cm/GW, respectively. An all-optical diode based on the 2D van der Waal NbC/GaS heterostructure achieves the nonreciprocity factor of 1.73 dB and transmittance symmetry of 0.81. The simulations with and without interfacial effect suggest the nonreciprocal nonlinear absorption and interfacial charge transfer in 2D van der Waals heterostructures responsible for the all-optical diode action. The 2D van der Waals heterostructures have the designable and excellent nonlinear optical properties, which is also integral, simple, and compatible with the complementary metal oxide semiconductor (CMOS) technology for all-optical diodes in optical signal processing.

## 2. Results and Discussion

A straightforward approach to design all-optical diode is to introduce axial asymmetry propagation, i.e., nonreciprocal transmission of light. This nonreciprocal transmission can be achieved in nonlinear optical regime owing to the permutation with opposite nonlinear absorption of SA and RSA materials. **Figure 1(a)** sketches the van der Waals heterostructure composed of SA and RSA materials, which exhibits the nonreciprocal transmission in nonlinearity due to the axially nonlinear asymmetry. When the light transmits from SA to RSA materials, the transmittance increases with the pump intensity, mimicking forward bias in electric diodes. However, the light transmits from RSA to SA materials, the transmittance decreases with the pump intensity, mimicking the reverse bias in electric diodes.

In addition to the RSA and SA responses, the interfacial charge transfer in van der Waals heterostructures is also considered due to the efficient charge transfer[10] and enhanced nonlinear optical properties in two-dimensional materials[21]. The charge transfer process in van der Waals heterostructures is displayed in **Figure 1(b)**. The interface effect on the all-optical diode can depend on $\sigma_{TPA}$, $\sigma_{12}$, $\sigma_s$, and $\sigma_e$, where $\sigma_{TPA}$ ($\sigma_{12}$) are two-photon (excited state) absorption cross-sections of RSA material and $\sigma_s$ ($\sigma_e$) are ground (excited) state absorption cross-sections of SA material.

To clarify the effects of SA, RSA, and interface in van der Waals heterostructures on the nonreciprocity and transmittance symmetry of all-optical diode, the fourth-order Runge-Kutta method (details in Supporting Information) was used. This numerical method can be used to simulate all-optical diode responses by solving light propagation equations through SA material, the interface, and RSA material. Corresponding propagation equations in SA and RSA materials are as follows.

$$\frac{dI}{dz'} = -\alpha_0^{SA} I - \beta_{SA} I^2 \tag{1}$$

$$\frac{dI}{dz'} = -\alpha_0^{RSA} I - \beta_{RSA} I^2 \tag{2}$$

Where, $\alpha_0^{SA}$ ($\alpha_0^{RSA}$) and $\beta_{SA}$ ($\beta_{RSA}$) are linear and nonlinear absorption coefficients of SA (RSA) material, respectively.

Accordingly, the light propagation through the interface region, which induces the charge transfer process, can be derived from rate equations (S25-S30 in Supporting Information) and can been expressed as:

$$\frac{dI}{dz'} = -N \frac{\left[\sigma_{TPA} I + \sigma_{12} \frac{I^2}{I_1} + \sigma_s \left(\frac{I^2}{I_2} - \frac{I^2}{I_1}\frac{\tau_3}{\tau_1}\right) + \sigma_e \left(\frac{I}{I_3} + \frac{\tau_4}{\tau_3}\right)\left(\frac{I^2}{I_2} - \frac{I^2}{I_1}\frac{\tau_3}{\tau_1}\right)\right]}{\left[1 + \frac{I^2}{I_1} + \left(\frac{I^2}{I_2} - \frac{I^2}{I_1}\frac{\tau_3}{\tau_1}\right) + \left(\frac{I}{I_3} + \frac{\tau_4}{\tau_3}\right)\left(\frac{I^2}{I_2} - \frac{I^2}{I_1}\frac{\tau_3}{\tau_1}\right)\right]} I \tag{3}$$

To describe the performance of all-optical diodes, the nonreciprocity factor[4] is defined as $F = 10 \times \log_{10}(T_{forward}/T_{reverse})$, where $T_{forward}$ and $T_{reverse}$ are the normalized transmittance in forward bias and reverse bias, respectively. The larger $F$ value means the better performance of all-optical diodes. Besides, transmittance symmetry of all-optical diodes about the axis $T=1$ can be characterized by the transmittance symmetry factor defined as $S = (T_{forward} - 1)/(1 - T_{reverse})$, which can serve as a criterion for evaluating the variations of $T_{forward}$ and $T_{reverse}$ to describe the on-off performance. When $S$ is less than 0, the SA/RSA material does not function as a diode. When *S* is close to 1, the diode demonstrates the perfect on-off performance.

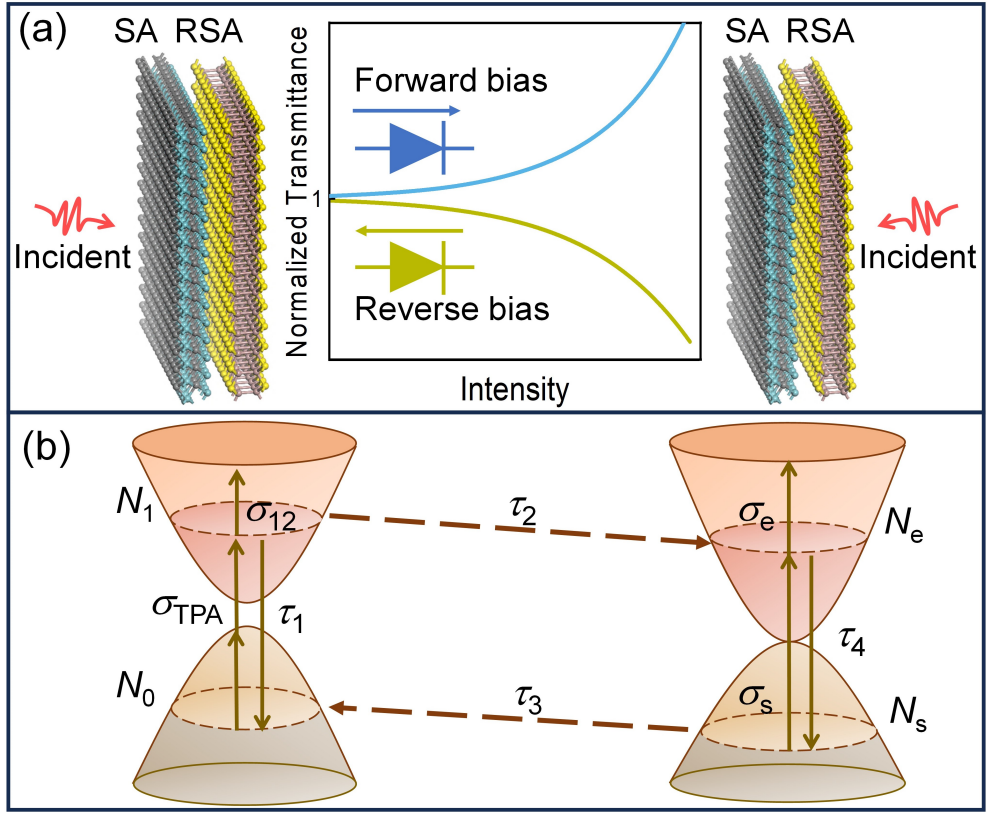

**Figure 1.** (a) Schematic of all-optical diode with forward bias and reverse bias; (b) charge transfer process at the interface of van der Waals heterostructures.

**Figure 2(a)** exhibits all-optical diode performance as a function of intensity with different $\alpha_0^{SA}$ values ($3\times10^4$ to $8\times10^4$ cm$^{-1}$). The corresponding *F* and *S* versus $\alpha_0^{SA}$ are displayed in **Figure 2(b)**. The *F* exhibits an increasing trend with the increase of $\alpha_0^{SA}$, which is due to large $\alpha_0^{SA}$ value contributing to large transmittance in forward bias and small transmittance in reverse bias. But the *S* almost keeps a constant value (~ 1) with the increase of $\alpha_0^{SA}$, which is due to the reason that *S* is mainly determined by nonlinear optical absorption ($\beta_{SA}$ and $\beta_{RSA}$) instead of linear optical absorption ($\alpha_0^{SA}$ and $\alpha_0^{RSA}$). Similarly, all-optical diode performances as a function of intensity with different $\alpha_0^{RSA}$ values (from $1.0\times10^4$ to $7.7\times10^4$ cm$^{-1}$) are also simulated as shown in **Figure 2(c)**. The $\alpha_0^{RSA}$-dependent *F* and *S* (**Figure 2(d)**) exhibit the same trend as those of $\alpha_0^{SA}$ as displayed in **Figure 2(c)**. These results demonstrate that the $\alpha_0^{SA}$ and $\alpha_0^{RSA}$ determined nonreciprocity performance of the all-optical diode, while they don't induce the variation of the symmetry. By a comparison, $\alpha_0^{SA}$-dependent $F$ illustrates a larger slope than that of $\alpha_0^{RSA}$-dependent $F$ as the thicker thickness of SA material than that of RSA material in simulation. Although the linear absorption of SA and RSA materials has a significant effect on the nonreciprocity, there is almost no effect on the transmittance symmetry.

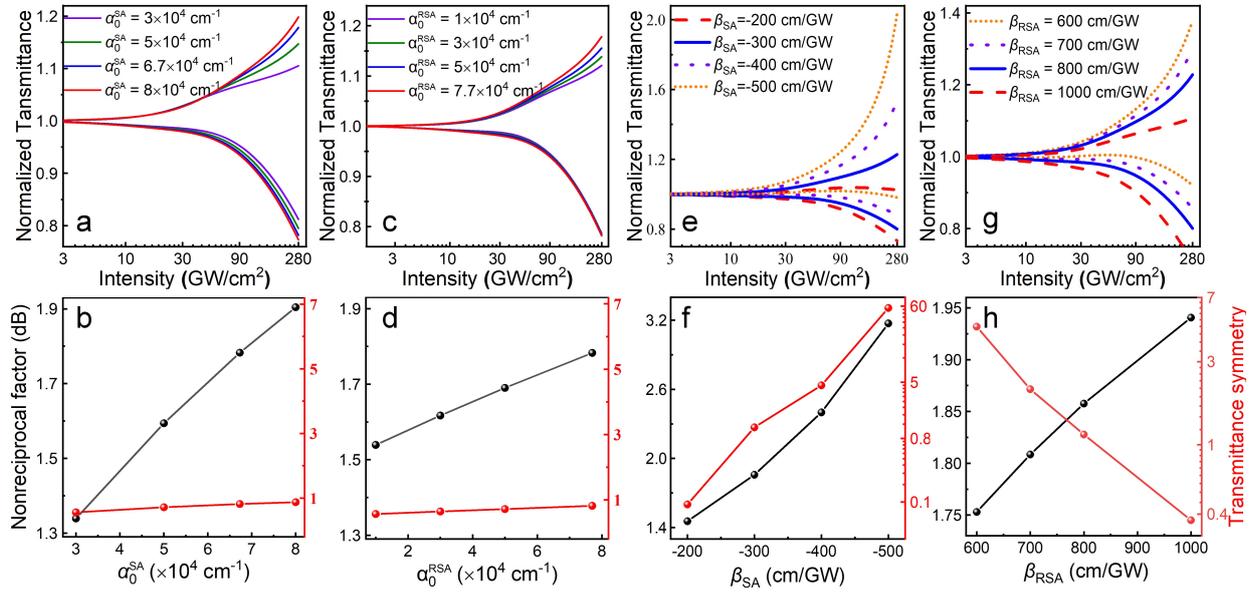

**Figure 2.** Simulation results of all-optical diode with different (a) $\alpha_0^{SA}$, (c) $\alpha_0^{RSA}$, (e) $\beta_{SA}$, and (g) $\beta_{RSA}$; nonreciprocal factor ($F$, black) and transmittance symmetry ($S$, red) versus (b) $\alpha_0^{SA}$, (d) $\alpha_0^{RSA}$, (f) $\beta_{SA}$, and (h) $\beta_{RSA}$. (In addition to the variable parameters, the other invariants are $\alpha_0^{SA} = 6.7 \times 10^4$ cm$^{-1}$, $\alpha_0^{RSA} = 7.7 \times 10^4$ cm$^{-1}$, $\beta_{SA} = -300$ cm/GW, $\sigma_{TPA} = 2.3 \times 10^{-18}$ cm$^4$/GW, $\sigma_{12} = 1.7 \times 10^{-18}$ cm$^2$, $\sigma_s = 1.0 \times 10^{-16}$ cm$^2$, and $\sigma_e = 1.4 \times 10^{-18}$ cm$^2$.)

To further explore the all-optical diode performances about transmittance symmetry, the simulations with different $\beta_{SA}$ and $\beta_{RSA}$ values are also performed. **Figure 2(e)** exhibits all-optical diode performances as a function of intensity with $\beta_{SA}$ values from -500 to -200 cm/GW, which shows obvious variations of $\beta_{SA}$-dependent $F$ and $S$. To demonstrate clearly, the $\beta_{SA}$-dependent $F$ and $S$ are calculated and depicted in **Figure 2(f)**. It is evident that the $F$ increases from 1.4 to 3.2 dB and $S$ increases from 0.1 to 56.4 when $\beta_{SA}$ varies from -200 to -500 cm/GW. When the $\beta_{SA}$ is close to -300 cm/GW, $S$ is close to 1.

Similarly, all-optical diode performances as a function of intensity with $\beta_{RSA}$ values from 600 to 1000 cm/GW are also simulated as shown in **Figure 2(g)**. The $F$ increases from 1.75 dB to 1.9 dB and $S$ decreases from 4.8 to 0.3 as the $\beta_{RSA}$ increases from 600 to 1000 cm/GW as depicted in **Figure 2(h)**. The increase of $F$ with $\beta_{SA}$ and $\beta_{RSA}$ suggests that the nonreciprocity is also related to the nonlinear optical absorption. Correspondingly, the symmetry $S$ is mainly determined by the nonlinear optical absorption instead of the linear optical absorption. In addition, the $\beta_{SA}$ induced $F$ has a bigger variation range than the $\beta_{RSA}$ induced $F$ due to the thicker thickness of SA material than that of RSA material in our simulation. This simulation affords a roadmap for the design and optimization of all-optical diode for better performance with better $\beta_{SA}$ and $\beta_{RSA}$ values.

Besides, the all-optical diode responses with different $\sigma_{TPA}$, $\sigma_{12}$, $\sigma_{s}$, and $\sigma_{e}$ values are also simulated to reveal the effect of interfacial charge transfer as shown in **Figure 3**. The all-optical diode with $\sigma_{TPA}$ values from $2.3 \times 10^{-19}$ to $5.7 \times 10^{-17}$ cm$^4$/GW are exhibited in **Figure 3(a)**, which suggests the same diode performance with different $\sigma_{TPA}$ values. Corresponding $F$ (1.85 dB) and $S$ (1.15) values are calculated and displayed in **Figure 3(b)**, which also suggest the F and S is insensitive of the nonlinear absorption at the interface. Similarly, the $\sigma_{12}$-dependent all-optical diode also simulated as shown in **Figure 3(c)** with the $\sigma_{12}$ from $1.7 \times 10^{-19}$ to $1.7 \times 10^{-16}$ cm$^2$. The $\sigma_{12}$-dependent $F$ and $S$ is also depicted in **Figure 3(d)**, which show almost no effect of $\sigma_{12}$ at the interface.

Furthermore, the all-optical diode with $\sigma_s$ values from $1.1 \times 10^{-17}$ to $1.0 \times 10^{-16}$ cm$^2$ are also simulated as exhibited in **Figure 3(e)** with the $F$ and $S$ versus $\sigma_s$ depicted in **Figure 3(f)**. It is evident that the $F$ and $S$ have a slight trend to rise with $\sigma_s$, which suggests the slight effect of $\sigma_s$ at the interface. As for the $\sigma_e$, there is an obvious effect on diode performance as shown in **Figure 3(g)** with the $\sigma_e$ from $1.4\times10^{-19}$ to $2.3\times10^{-17}$ cm$^2$. As shown in **Figure 3(h)**, the $F$ and $S$ demonstrates the obviously increased trend for $F$ (1.7-3.3 dB) and $S$ (1.1-1.7) with the $\sigma_e$. The effect of $\sigma_{TPA}$, $\sigma_{12}$, $\sigma_s$, and $\sigma_e$ on the diode performance at the interface can be understood by Equation (3), which is non-negligible when designing the all-optical diode.

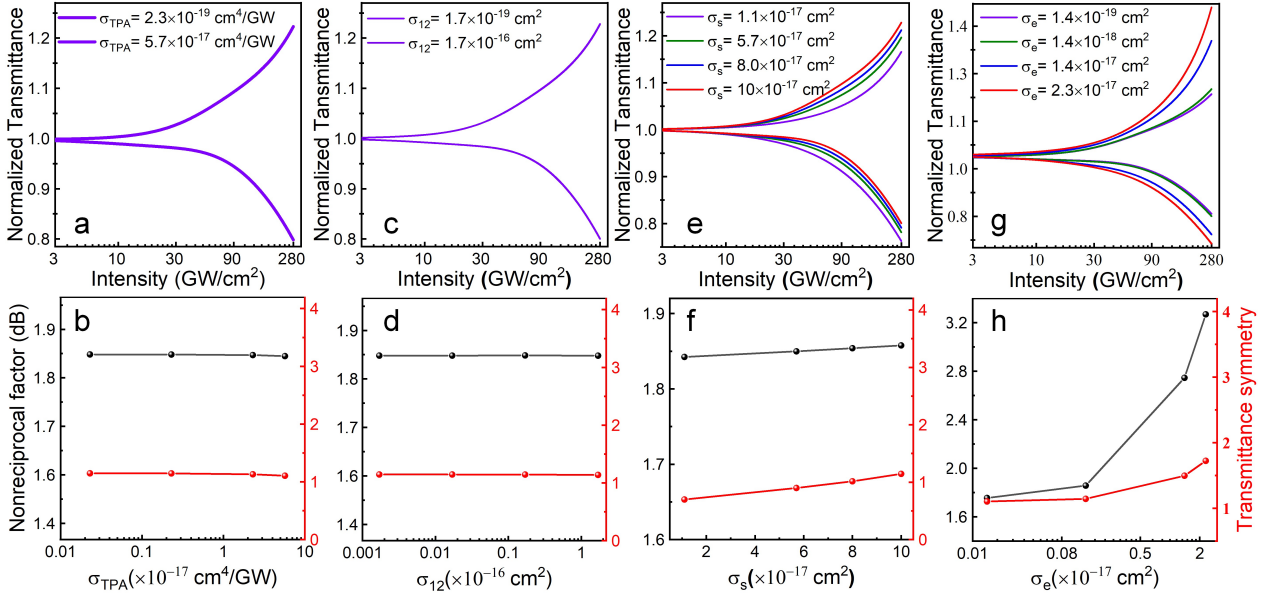

**Figure 3.** Simulation results of all-optical diode with (a) $\sigma_{TPA}$, (c) $\sigma_{12}$, (e) $\sigma_s$, and (g) $\sigma_e$; nonreciprocal factor ($F$, black) and transmittance symmetry ($S$, red) versus (b) $\sigma_{TPA}$, (d) $\sigma_{12}$, (f) $\sigma_s$, and (h) $\sigma_e$.

According to the simulation in **Figure 2** and **3**, the nonlinear absorption properties of SA and RSA materials play the vital role in the fabrication of all-optical diode. Recently reported MXene materials exhibit the excellent SA property[15], which is good for the all-optical diode. However, the functional groups[16] in MXene have the strong effect on the SA coefficients of the MXene, which is uncontrollable. Herein, NbC is chosen as the SA materials with the high stability to avoid the uncontrollability. NbC film is prepared via the liquid phase exfoliation method as illustrated in **Figure S1(a)** (Supporting Information) and characterized by the X-ray diffraction (XRD) (**Figure S2(a)** in Supporting Information) to demonstrate the successful preparation. To measure the $\beta_{SA}$, the open-aperture Z-scan system excited by the 800-nm laser with the pulse duration of 150 fs and the repetition rate of 1 kHz is employed. **Figure 4(a)** displays the Z-scan curves of NbC film under different on-focus intensities (the intensity at focal point of lens). The normalized optical transmittance increases with the optical intensity increasing and approaches the maximum when the sample position *z* approaching the focal point (*z* = 0 mm) of the lens (focal length: 17.5 mm). This result exhibits the typically SA behavior, which is induced by Pauli-blocking effect. In detail, due to the metal property[22] of NbC, the single-photon absorption from valence band to conduction band can occurred easily. Under high enough optical intensity, all possible states in conduction band can be occupied, which causes the SA response.

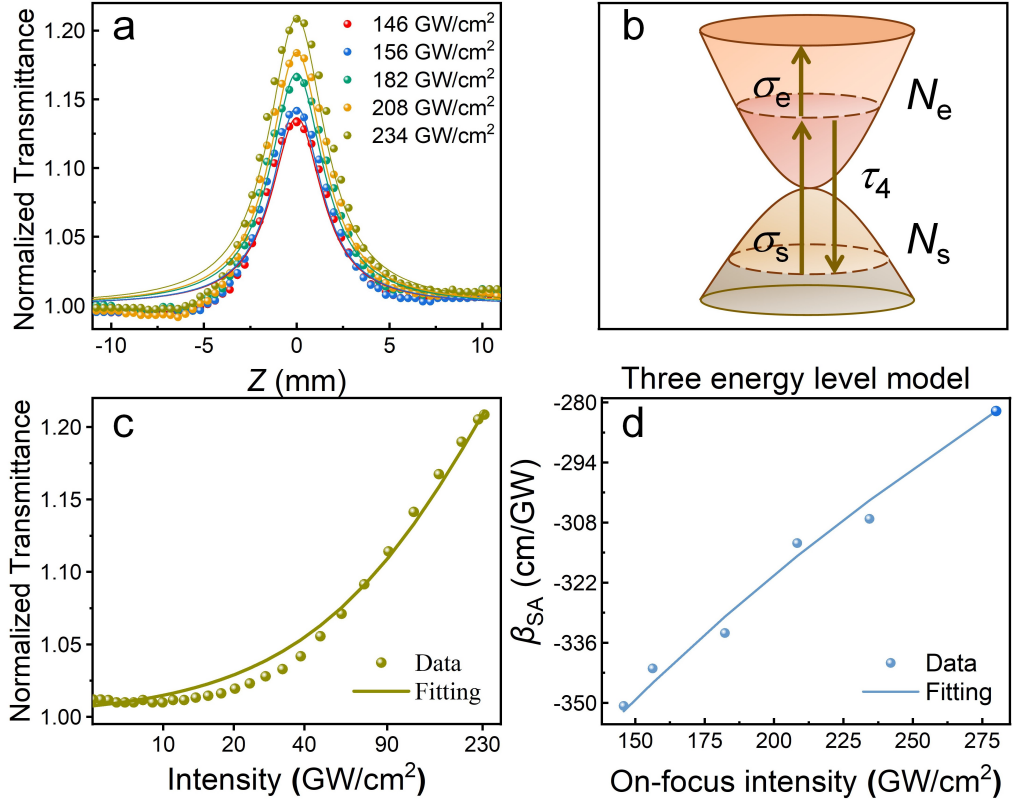

**Figure 4.** (a) Open-aperture Z-scan traces of NbC film under different on-focus intensity excitation, (b) three energy level model, (c) normalized optical transmittance as a function of $I_z$, and (d) nonlinear absorptive coefficient of NbC film versus on-focus intensity.

To accurately clarify the physical mechanism of the SA occurring in NbC film, the single-photon absorption energy-level model in **Figure 4(b)** was established. According to the corresponding rate equations (S1-S6 in Supporting Information), the analytic expression of transmittance, $T(I_z)$ versus optical intensity is solved, which can be described as following:

$$T_{SA}(I_z) = \left(1 - M \frac{1 + \kappa \frac{I_z}{I_s}}{1 + \frac{I_z}{I_s}}\right) / (1 - M) \tag{4}$$

Where $I_z$ is optical intensity in different position $z$; $I_s$ is the saturable intensity; $T(I_z)$ is normalized transmittance as a function of $I_z$; $\kappa$ is the ration of excited state absorption cross-section ($\sigma_e$) and ground state cross-section ($\sigma_s$); and $M$ is equal to $\alpha_0^{SA} L_{SA}$, where $\alpha_0^{SA}$ is linear absorption coefficient of NbC extracted from **Figure S2(b)** (Supporting Information) and $L_{SA}$ is the thickness of NbC film measured by atomic force microscope (AFM) as shown in **Figure S2(e)** (Supporting Information).

To extract the $\kappa$ value by fitting the Z-scan data using Equation (4), the Z-scan data in **Figure 4(a)** under excitation intensity of 234 GW/cm² are transferred into $I_z$-dependent normalized transmittance (dots in **Figure 4(c)**) by using equation $I_z = I_0 / (1 + z^2/z_0^2)$, where $I_0$ is the on-focus intensity and $z_0 = \pi \omega_0^2 / \lambda$ ($\omega_0$ is the waist radius of the pump light). The optical parameters of $\kappa$ (~0.28) and $I_s$ (154 GW/cm²) are extracted from the fitting in **Figure 4(c)**, which corresponds well with the experimental data. Similarly, Z-scan data under excitation of different on-focus intensities of 146, 156, 182, and 208 GW/cm² in **Figure** 4(a) are also fitted by Equation (4) as shown in **Figure S3** (Supporting Information). All extracted $\kappa$ values shown in Table S1 (Supporting Information) are less than 1, which reveals the large ground state cross-section ($\sigma_s$) than the excitation state cross-section ($\sigma_e$). The smaller $\kappa$ value means the stronger SA ability. As such, the decreased $\kappa$ value with the increase of the on-focus intensity demonstrates the enhanced SA response in **Figure 4(a)**. In addition, the extracted $I_s$ values under the different on-focus intensities remain almost the same value around 150 GW/cm². This $I_s$ value of NbC is larger than those of Ti$_3$C$_2$T$_x$[15], MOF[23], Nb$_2$C[24], SnS[25], and MoSe$_2$[26], which implies the high damage threshold benefiting for nonlinear high-power photonic devices.

Note that the nonlinear SA coefficients as an important parameter for designing all-optical diode, were extracted by fitting the Z-scan data in **Figure 4(a)** using the following Equation (5)[27]:

$$T(z) = \sum_{m=0}^{3} \frac{\left[ -\beta I_0 L_{\text{eff}} / \left(1 + z^2/z_0^2\right) \right]^m}{(m+1)^{3/2}} \qquad (5)$$

Where $\beta$ is the nonlinear absorption coefficient and $T(z)$ is the normalized transmittance as the function of the sample position $z$. The $L_{\text{eff}} = (1-e^{-\alpha_0 L})/\alpha_0$ is the effective thickness of the sample ($L$: thickness of the sample; $\alpha_0$: linear absorption coefficient of the sample).

**Figure 4(d)** depicts the value of $\beta_{\text{SA}}$ varying from -350 to -300 cm/GW under on-focus intensities of 146-234 GW/cm$^2$ and is summarized in Table S1 in Supporting Information. The experimental $\beta_{\text{SA}}$ value is close to the simulated value in **Figure 2**, which suggests the NbC as an ideal SA material for all-optical diode. Besides, the $\beta_{\text{SA}}$ of NbC is 2-5 orders in magnitude larger than those of germanium[28], Ti$_3$C$_2$T$_x$[15], MOF[23], Nb$_2$C[24], tellurene[29], GeSe[6], SnS[25], MoSe$_2$[26], and MoS$_2$[30] as shown in Table 1. More intriguingly, the $\beta_{\text{SA}}$ values exhibit the increased trend with the on-focus intensity. This tendency can be well fitted by the equation (S10 in Supporting Information) as shown in **Figure 4(d)**. This fitting result also gives the $I_s$ value of ~148 GW/cm$^2$, which is consistent to the $I_s$ value extracted from Equation (4). Furthermore, the imaginary part of third-order polarizability (Im $\chi^3$) is related to $\beta_{\text{SA}}$, which can be calculated by Equation (6)[15]:

$$\text{Im}\,\chi^3\,(esu) = \left[\frac{10^{-7} c\lambda n^2}{96\pi^2}\right]\beta \qquad (6)$$

Where $c$ is the velocity of light, $\lambda$ is the wavelength of the pump laser, and $n$ is the real refractive index of the sample. The large value of Im $\chi^3$ (-1.52 × 10$^{-8}$ esu in Table S1 in Supporting Information) suggests the high nonlinear optical response of NbC sample.

To eliminate the effect of linear absorption coefficient on Im $\chi^3$, the figure of merit ($\text{FOM} = \left|\text{Im}\,\chi^3/\alpha_0\right|$) is calculated to characterize the nonlinear response of the sample. As a comparison, the absolute value of Im $\chi^3$ and FOM of NbC exhibit the larger values than the other two-dimensional materials such as germanium[28], Ti$_3$C$_2$T$_x$[15], MOF[23], tellurene[29], GeSe[6], SnS[25], MoSe$_2$[26], and MoS$_2$[30] as shown in Table 1, indicating excellent nonlinear optical performance for potential applications in all-optical diodes.

**Table 1.** Optical parameters of NbC film and other 2D materials under the excitation

wavelength of 800 nm.

| Materials | Laser parameters | $\beta_{SA}$ (cm/GW) | Im $\chi^3$ (esu) | FOM (esu cm) | $I_s$ (GW/cm$^2$) | Ref. |
|---|---|---|---|---|---|---|
| NbC | 150 fs/1 kHz | -350 | -1.52×10$^{-8}$ | 2.27×10$^{-13}$ | 150 | This work |
| Ge | 35 fs/2 kHz | -0.06 | -3.03×10$^{-13}$ | 9.6×10$^{-15}$ | - | [28] |
| Ti$_3$C$_2$T$_x$ | 35 fs/2 kHz | -0.29 | -2.50×10$^{-13}$ | - | 89 | [15] |
| MOF | 100 fs/1 kHz | -0.03 | -2.00×10$^{-14}$ | - | 30 | [23] |
| Nb$_2$C | 35 fs/2 kHz | -0.28 | -3.70×10$^{-11}$ | - | 99 | [24] |
| Tellurene | 35 fs/2 kHz | -0.11 | -1.11×10$^{-13}$ | 1.22×10$^{-14}$ | - | [29] |
| GeSe | 56 fs/1 kHz | -0.02 | -1.81×10$^{-14}$ | 3.64×10$^{-15}$ | - | [6] |
| SnS | 100 fs/1 kHz | -0.05 | -3.05×10$^{-13}$ | - | 35 | [25] |
| MoSe$_2$ | 226 fs/15 kHz | -1.10 | -7.72×10$^{-9}$ | -3.36×10$^{-14}$ | 112 | [26] |
| MoS$_2$ | 100 fs/1 kHz | -0.006 | -3.30×10$^{-15}$ | 1.16×10$^{-15}$ | - | [30] |

As for the RSA material, a 2D material with a large bandgap is chosen to exhibit the RSA response under the excitation wavelength of 800 nm (1.55 eV). The GaS presents a bandgap of 2.29 eV obtained from **Figure S2(c)** (Supporting Information), which can be excited below the bandgap when interaction with the 800 nm light. GaS films are prepared by the physical vapor deposition method as shown in **Figure S1(b)** (Supporting Information) and characterized by XRD (**Figure S2(a)** in Supporting Information). The $\beta_{RSA}$ of the GaS film was also measured in the same open-aperture Z-scan system under the excitation of 800 nm femtosecond laser. As shown in **Figure** 5(a), the Z-scan experimental results of the GaS film under different on-focus intensities exhibit the decreased normalized transmittance when the sample approaches the focal point of the lens. This phenomenon is the typical RSA response. To clarify the mechanism of the RSA in GaS film, the relationship of ln(1-$T$) versus ln($I_z$) are depicted in **Figure 5(b)** and fitted linearly, in which $T$ is the normalized transmittance. The slope value of 1 in the ln(1-$T$) vs. ln($I_z$) curve indicates the that the RSA is induced by two-photon absorption (TPA) response[31]. However, the slope value of 1.22 is observed in **Figure 5(b)**, which is attributed to the contribution of excited state absorption to RSA in addition to TPA.

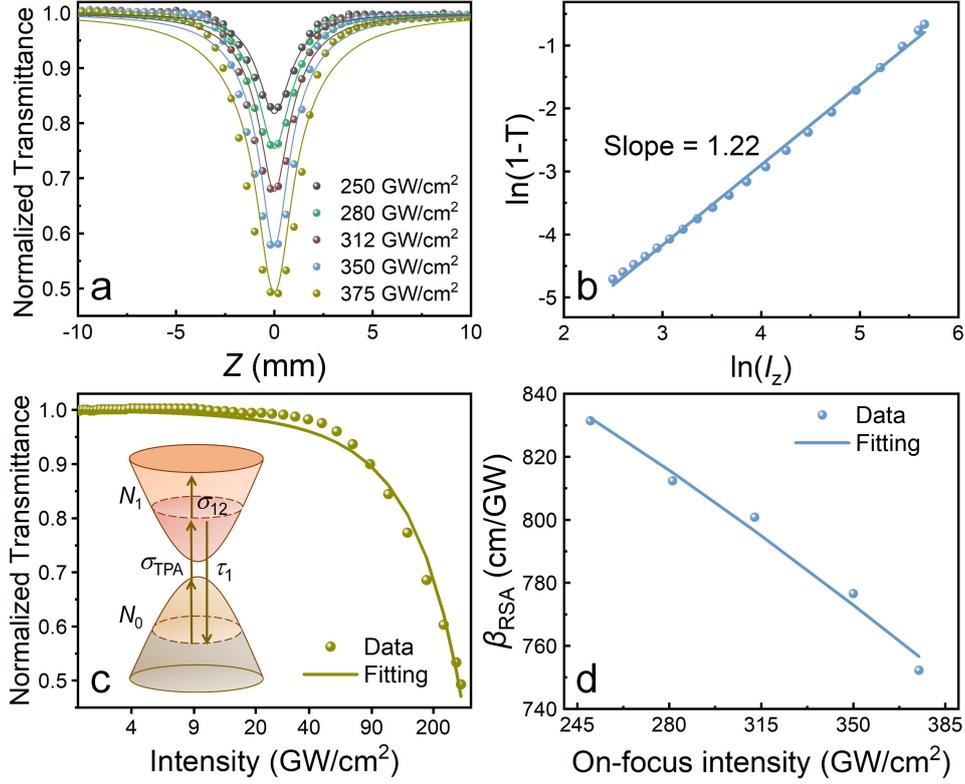

**Figure 5.** (a) Open-aperture Z-scan results of GaS film under different on-focus intensities, (b) plot of ln(1-T) vs. ln($I_z$), (c) normalized optical transmittance (inset is TPA energy-level model), and (d) TPA coefficient of GaS film versus optical intensity.

Because the optical bandgap (2.29 eV) of the GaS film is larger than the photon energy (1.55 eV) of the excitation light, GaS sample needs absorb two photons to excite the electrons in valence band into conduction band. The TPA energy-level model as shown in the inset of **Figure 5(c)** is established, which is used to explain the physical mechanism of RSA in GaS film by Equation 7 (S18 in Supporting Information):

$$T_{\mathrm{RSA}}(I_z) = \left(1 - N_{\mathrm{RSA}} \frac{\left(\sigma_{\mathrm{TPA}} I_z + \sigma_{12} \frac{I_z^2}{I_{\mathrm{s,2pa}}^2}\right)}{\left(1 + \frac{I_z^2}{I_{\mathrm{s,2pa}}^2}\right)}\right) \Big/ \left(1 - \alpha_0^{\mathrm{RSA}} L_{\mathrm{RSA}}\right) \quad (7)$$

Where $\sigma_{\mathrm{TPA}}$ is TPA cross-section; $\sigma_{12}$ is excited state absorption cross-section; $\tau_{10}$ is relaxation time; $N_{\mathrm{RSA}}$ is molecular density of GaS; $\alpha_0^{\mathrm{RSA}}$ is linear absorption coefficient of GaS extracted from **Figure S2(b)** (Supporting Information) and $L_{\mathrm{RSA}}$ is the thickness of GaS film measured by AFM as shown in **Figure S2(d)** (Supporting Information).

To better understand the RSA response of GaS film, the experiment data of GaS film in **Figure 5(a)** are transferred into the $I_z$-dependent normalized transmittance under the on-focus intensity of 375 GW/cm$^2$, which is fitted by Equation (7) as shown in **Figure 5(c)**. The fitting result gives the $\sigma_{TPA}$ value (7.65×10$^{-27}$ cm$^4$/GW) and $\sigma_{12}$ value (1.9×10$^{-18}$ cm$^2$), which further demonstrates that in addition to the TPA response, the excited state absorption also has an effect on the RSA in GaS film. To illuminate to what extent the excited state absorption affects the RSA compared with the TPA, the GaS film under on-focus intensities of 250, 280, 312, and 350 GW/cm$^2$ are also fitted by Equation (7) (**Figure S4** in Supporting Information) and the $\sigma_{TPA}$ and $\sigma_{12}$ values are summarized in Table S3 (Supporting Information). The $\sigma_{TPA}/\sigma_{12}$ values exhibit the increase trend with the on-focus intensity increase, which reveals that the increased contribution of TPA and decreased contribution of excited state absorption to RSA as on-focus intensity increases.

Equation (5) was also utilized to fit the Z-scan data in **Figure 5(a)** to extract the nonlinear absorption coefficients, i.e., TPA coefficients of the GaS film. The extracted $\beta_{RSA}$ value varies from 831 to 752 cm/GW (Table S2 in Supporting Information) under the on-focus intensities of 250-375 GW/cm$^2$, which is desirable for all-optical diode according to the simulation in **Figure 2**. The $\beta_{RSA}$ value of GaS is larger than that of Ti$_3$C$_2$T$_x$[32], TlGaS$_2$[33], and some transition-metal dichalcogenide materials[19, 34-36], revealing excellent nonlinear absorption ability for GaS as a potential RSA material. The $\beta_{RSA}$ value also exhibits the decreasing tendency with the on-focus intensity as shown in **Figure 5(d)**, which can be well fitted by Equation S22 (Supporting Information). The $I_{s,2pa}$ of 296 GW/cm$^2$ in GaS is larger than that in MoS$_2$ (64.5 GW/cm$^2$) [18], which means high damage threshold for nonlinear photonic devices with GaS. Besides, Im $\chi^3$ and FOM of GaS film are also calculated under different intensities as summarized in Table S2 (Supporting Information). The larger Im $\chi^3$ and FOM values than the other 2D materials such as Ti$_3$C$_2$T$_x$[32], TlGaS$_2$[33], MoS$_2$[34], WS$_2$[19], SnS$_2$[36], and SnSe$_2$[36] are shown in Table 2, further suggests GaS as the excellent nonlinear material for all-optical diodes.

**Table 2.** TPA optical parameters of GaS film and the other 2D materials.

| Materials | Laser parameters | $\beta_{RSA}$ (cm/GW) | Im $\chi^3$ (esu) | FOM (esu cm) | Ref. |
|---|---|---|---|---|---|
| GaS | 150 fs/1 kHz | 931 | $1.1\times10^{-7}$ | $1.42\times10^{-12}$ | This work |
| MoS$_2$ | 50 fs/1 kHz | 0.049 | $1.4\times10^{-12}$ | $7\times10^{-12}$ | [34] |
| Ti$_3$C$_2$T$_x$ | 100 fs/100 kHz | 0.014 | $2.45\times10^{-14}$ | $4\times10^{-15}$ | [32] |
| WS$_2$ | 40 fs/1 kHz | 525 | $2.72\times10^{-9}$ | $2.51\times10^{-15}$ | [19] |
| TlGaS$_2$ | 300 fs/1 kHz | 0.74 | $9.08\times10^{-11}$ | - | [33] |
| TiS$_2$ | 7 ns/10 Hz | 62 | - | - | [35] |
| SnS$_2$ | 21 ps/~ | 0.3 | $9.36\times10^{-14}$ | $2.05\times10^{-13}$ | [36] |
| SnSe$_2$ | 21 ps/~ | 0.038 | $1.19\times10^{-14}$ | $3.6\times10^{-14}$ | [36] |

To verify the performance of all-optical diode of NbC/GaS heterostructure, the Z-scan data under the forward and reverse incidence is measured and shown in **Figure 6(a)** under the pump intensity of 280 GW/cm$^2$. It is evident that under the forward incidence, the NbC/GaS heterostructure demonstrates the SA response, while under the reverse incidence, the heterostructure demonstrates the RSA response. The intensity dependent normalized transmittance under both forward and reverse incidence is displayed in **Figure 6(b)** with the typical all-optical diode behavior. When pump intensity is below 8.4 GW/cm$^2$, the NbC/GaS heterostructure exhibits mainly the linear absorption, which is analog to the knee-voltage in the electric diode. When the pump intensity increases further, the NbC/GaS heterostructure exhibits mainly the nonlinear absorption, mimicking the electronic diode response.

To understand the diode response of NbC/GaS heterostructure without the effect of interface, the fourth-order Runge-Kutta method was used to numerically simulate the diode response in forward and reverse incidences by successively solving Equations (1-2) in NbC and GaS with the experimental values of $\beta_{SA}$ = -282 cm/GW and $\beta_{RSA}$ = 812 cm/GW obtained in **Figure 4(d)** and **Figure 5(d)**. The simulation result is shown in **Figure 6(c)**, which catches the main features of all-optical diode in **Figure 6(b)** but with a poor transmittance symmetry. The notable difference regarding the transmittance symmetry between experiment and simulation is due to the interfacial charge transfer effect in the van der Waals NbC/GaS heterostructures. As such, Equation (3) with the interface effect is taken into accounted in the simulation as shown in **Figure 6(d)**. It is evident that both the experiment (**Figure 6(b)**) and the simulation (**Figure 6(d)**) matches well with each other, which exhibits the all-optical diode behavior with the same performances of $F$ (1.73 dB) and $S$ (0.81) as that of the experimental result in **Figure 6(b)**. This result suggests the all-optical diode based on van der Waals materials is affected not only by the nonreciprocal nonlinear absorption of NbC and GaS films but also by the interfacial charge transfer at the interface. The interfacial charge transfer can result in the variations of absorption cross-section parameters of $\sigma_{TPA}$, $\sigma_{12}$, $\sigma_s$, and $\sigma_e$ as discussed in performance, while the $\sigma_s$ also has a slight effect. The larger $\sigma_s$ and $\sigma_e$ value, the greater impact of the interfacial charge transfer on the diode performance. The $\sigma_s$ of $1.2\times10^{-16}$ cm$^2$ and $\sigma_e$ of $1.6\times10^{-18}$ cm$^2$ at the interface of van der Waals NbC/GaS heterostructures are the best values in the simulation in **Figure 6(d)**. Besides, the NbC/GaS diode demonstrates the better balance between the transmittance symmetry and the nonreciprocity compared with that in graphene/C$_{60}$ [5]. More importantly, the results also suggests that the comprehensive physical mechanisms involving both nonreciprocal nonlinear absorption and interfacial charge transfer are responsible for all-optical van der Waals based diode analysis, which paves the way for all-optical communication and network in the future.

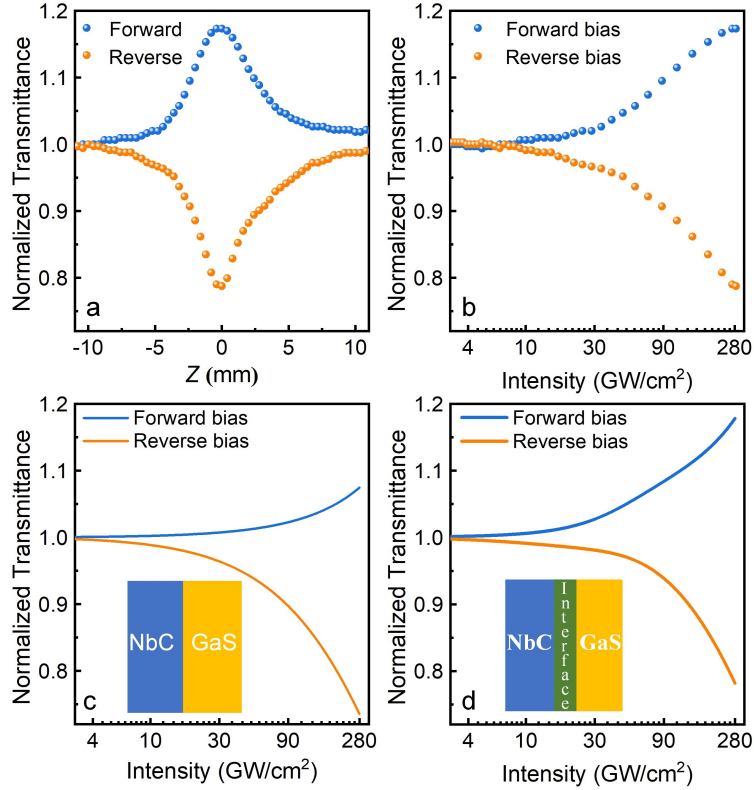

**Figure 6.** (a) Z-scan curves of NbC/GaS heterostructures in forward (blue) and reverse (orange) biases, (b) all-optical diode response in forward (blue) and reverse (orange) biases with different pump intensity, (c) simulated all-optical diode response of NbC/GaS heterostructures without the effect of interface layer, and (d) simulated all-optical diode response of NbC/GaS heterostructures with the interfacial charge transfer.

**Figure 3**. According to the simulation of the interfacial effect in **Figure 3**, the $\sigma_e$ mainly affects the diode

## 3. Conclusion

In summary, we have successfully demonstrated the all-optical diode based on 2D van der Waals NbC/GaS heterostructure in both experiment and simulation. Our study reveals the critical factors for fabricating and understanding all-optical diodes based on 2D van der Waals materials: (i) large nonlinear absorption coefficients of both SA and RSA materials are crucial for achieving optical nonreciprocity; (ii) in addition to the nonreciprocal nonlinear absorption in 2D van der Waals heterostructures, the interfacial charge transfer also has the essential effect on the diode responses. A thorough comprehension for the physical mechanism of all-optical diodes will significantly enhance the convenience of their implementation in optical signal processing based on 2D van der Waals heterostructures.

**Acknowledgements**

This work was supported by the National Natural Science Foundation of China (No. 12261141662, 12074311).

**Supporting information.** Preparation and characterization of samples, three energy-level model for single-photon absorption in NbC film, normalized optical transmittance of NbC film versus $I_z$, relationship between nonlinear absorption coefficient of NbC and $I_0$, energy-level model for two-photon absorption (TPA) in GaS film, normalized optical transmittance of GaS film versus $I_z$, relationship between nonlinear absorption coefficient of GaS and $I_0$, numerical simulation of all-optical diode, and energy-level model for van der Waals heterostructure in NbC/GaS heterostructure.

# Supporting Information

**All-optical diode via nonreciprocal nonlinear absorption and interfacial charge transfer in two-dimensional van der Waals heterostructures**

*Erkang Li, Jinhong Liu, Yanqing Ge, Mingjian Shi, Yijie Wang, Chunhui Lu\*, Yixuan Zhou\*, Xinlong Xu\**

Shaanxi Joint Lab of Graphene, State Key Laboratory of Photon-Technology in Western China Energy, International Collaborative Center on Photoelectric Technology and Nano Functional Materials, Institute of Photonics & Photon-Technology, School of physics, Northwest University, Xi'an 710069, China

E-mail: yxzhou@nwu.edu.cn (Yixuan Zhou); luchunhui@nwu.edu.cn (Chunhui Lu); xlxuphy@nwu.edu.cn (Xinlong Xu)

1. **Preparation and characterization of samples**

The NbC film was prepared via a liquid-phase exfoliation (LPE) method as shown in Figure S1(a). The 200 mg NbC powder and 150 mL 40% alcohol were mixed together. To exfoliate the few-layer NbC from the NbC powder, the supersonic machine (Qsonica Q700) was used to sonicate the mixed solution for 90 minutes in a water bath. Subsequently, the sonicated NbC solution was centrifugated at 3000 rpm (revolutions per minute) for 10 minutes to obtain NbC nanosheets supernatant. Last but not least, the 30 mL supernatant was filtered by a vacuum filtration method to obtain NbC film.

The GaS film was prepared by a physical vapor deposition (PVD) method, a bottom-up synthesis, in a single temperature zone tube furnace under a low-pressure condition as shown in Figure S1(b). The 10 mg GaS precursor was placed at the center of the temperature zone. A *C*-plane sapphire substrate (1×1 cm$^2$) was put downstream to deposit the GaS films. Before the growth, the air in the tube was purged by argon (Ar, 99.999%) with a flow velocity of 40 sccm (standard cubic centimeter per minute) to avoid the effect of oxygen on the growth of GaS films. During the growth, the temperature of the tube furnace was heated to 940 °C within 30 minutes and maintained 40 minutes to evaporate GaS powders into GaS vapor. In this process, the GaS vapor was carried downstream by 40 sccm Ar gas and then formed films

In the end, the GaS/NbC heterostructure was prepared by combining PVD and LPE methods together. The NbC film, prepared by the LPE method, was transferred onto the GaS film synthesized by PVD method to form the NbC/GaS heterostructure as shown in Figure S1.

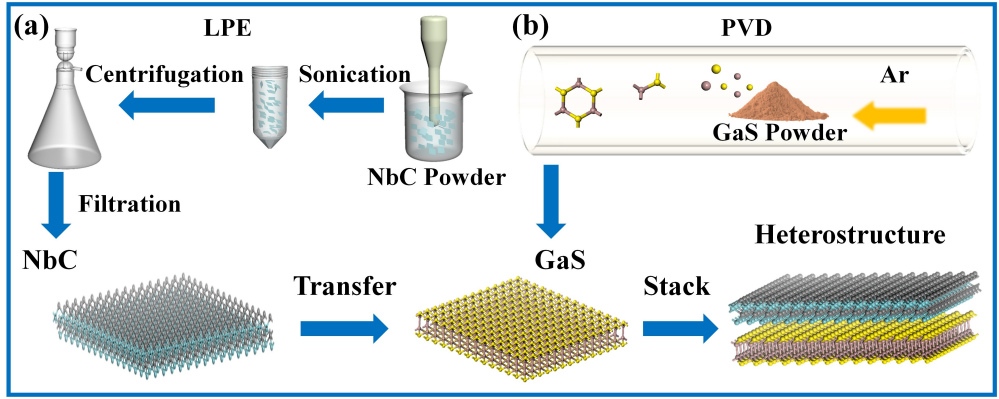

**Figure S1**. Schematic diagrams of preparation of GaS/NbC heterostructure and (a) NbC film via LPE method as well as (b) GaS film via PVD method.

To confirm the successful preparation of the samples, the X-ray diffraction (XRD, Thermo Fisher, ESCALAB Xi+) patterns of GaS, NbC, and GaS/NbC heterostructure were measured and displayed in Figure S2(a). The diffraction peaks of GaS film at 11.6°, 23.2°, 29.3°, and 35.0° corresponding to (002), (004), (101), and (006) lattice faces of hexagonal-structure GaS (PDF#08-0417) respectively, which are consistent with the previous report[1]. Similarly, the diffraction peaks of NbC film at 35.0°, 40.3°, 58.3°, 69.7°, 73.3°, and 87.1° are exhibited in Figure S2(a) corresponding to (111), (200), (220), (311), (222), and (400) lattice faces of cubic-structure NbC (PDF#38-1364), respectively. Figure S2(a) also displays the XRD patterns of GaS/NbC heterostructure encompassing all diffraciton peaks of GaS and NbC, demonstrating the successful preparation of GaS/NbC heterostructure. Here, the red diffraction peak in the top panel of Figure S2(a) is common in both GaS and NbC.

Figure S2(b) shows the linear optical absorption spectra of GaS and NbC films, which exhibits strong optical absorption response. According to the linear absorption spectrum, linear absorption coefficients of GaS ($\alpha_0^{RSA}$ = 77008 cm$^{-1}$) and NbC ($\alpha_0^{SA}$ = 67281 cm$^{-1}$) films can be calculated by using Lambert-Beer's law. In addition, based on the energy-dependent absorption in Figure S2(b), the optical bandgap ($E_g \sim 2.29$ eV) of GaS film can be determined by Tauc plot as shown in Figure S2(c). As for NbC, the gapless structure has been verified from electronic band structure[2].

The thicknesses of GaS and NbC films, which are the essential parameters for calculating the nonlinear optical absorption coefficient, were characterized by atomic force microscope (AFM, Bruker, Dimension Icon) as shown in Figure S2(d-e). The insets in Figure S2(d-e) accurately present the thicknesses of GaS film (~37 nm) and NbC film (~81 nm).

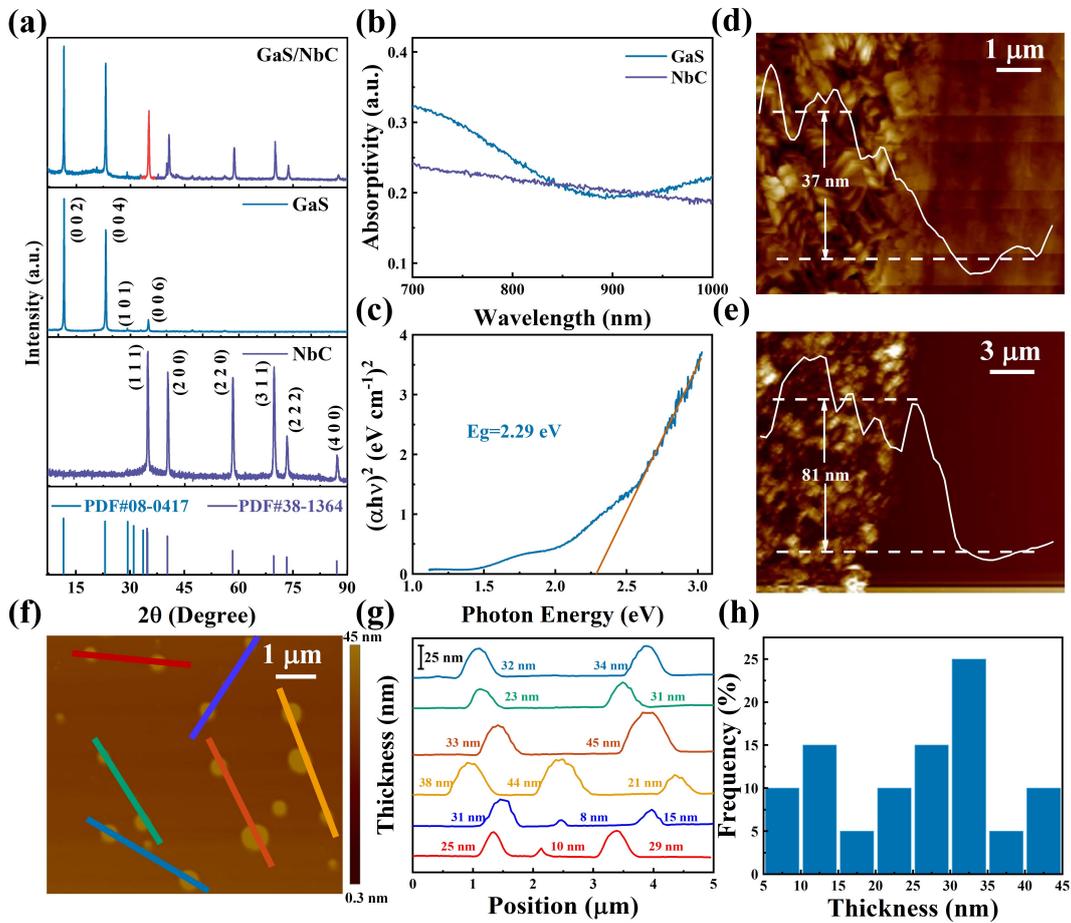

**Figure S2.** (a) XRD patterns of GaS/NbC heterostructure, GaS, NbC, and JCPDS PDF#08-0417 as well as PDF#38-1364; (b) UV-Vis spectra of GaS and NbC films; (c) Tauc plot of GaS film; AFM images of (d) GaS and (e) NbC films (Inset: thickness information); (f) AFM image, (g) height profiles, and (h) statistical thickness distribution of NbC nanosheets. (The a.u. stands for arbitrary units.)

To demonstrate the successful exfoliation of NbC nanosheets, AFM images were measured to acquire the thickness information as shown in Figure S2(f-g). Figure S2(f) shows the AFM images of NbC nanosheets in the range of $8 \times 8$ μm². Corresponding height profiles were extracted and presented in Figure S2(g). The lateral and longitudinal dimensions of NbC flakes are 400 nm and 25 nm in average, respectively. In addition, the thickness information was counted and displayed in Figure S2(h), which indicates the thickness of NbC nanosheets ranging from 5 nm to 45 nm.

## 2. Three energy-level model for single-photon absorption in NbC film

The rate equations of three energy-level model (Figure 4(b) in the main text) for the single-photon absorption in NbC film are as follows:

$$\frac{dN_s}{dt} = -\frac{N_s \sigma_s I}{h\nu} + \frac{N_e}{\tau} \tag{S1}$$

$$\frac{dN_e}{dt} = \frac{N_s \sigma_s I}{h\nu} - \frac{N_e}{\tau} \tag{S2}$$

$$N_{SA} = N_e + N_s \tag{S3}$$

$$\frac{dI}{dz'} = -\sigma_e N_e I - N_s \sigma_s I \tag{S4}$$

Where $N_e$ ($N_s$) and $\sigma_e$ ($\sigma_s$) are the carrier population densities and absorption-section in the excited state (ground state) of NbC; $N_{SA}$ is the total population densities; $\tau$ is the relaxation time of electrons from excited state to ground state; $I$ is the optical intensity and $z'$ is the light transmission position in the sample.

The Equation (S4) can also be described as following equation:

$$\frac{dI}{dz'} = -\alpha_{SA}(I) I \tag{S5}$$

Where $\alpha_{SA}(I)$ is the total absorption coefficient of NbC film.

Combining the Equation (S1-S5), the total absorption of NbC film under the steady state approximation can be solved and expressed as:

$$\alpha_{SA}(I) = \alpha_0^{SA} \frac{1 + \kappa \dfrac{I}{I_s}}{1 + \dfrac{I}{I_s}} \tag{S6}$$

Here, $\alpha_0^{SA}$ is linear absorption coefficient of NbC film, $\kappa = \sigma_e / \sigma_s$ and $I_s = (h\nu)/(\sigma_s \tau_1)$.

According to $T_{SA} = [1-\alpha_{SA}(I)L_{SA}]/[1-\alpha_0^{SA}L_{SA}]$ ($L_{SA}$ is the thickness of NbC film), the normalized transmittance of NbC film with the optical intensity at different sample position $z$ in Z-scan system can be expressed as:

$$T_{SA}(I_z) = \left(1 - M \frac{1 + \kappa \frac{I_z}{I_s}}{1 + \frac{I_z}{I_s}}\right) / (1 - M) \tag{S7}$$

Where $M$ ($\alpha_0^{SA} L_{SA}$) is the modulation depth, $I_s$ is the saturable intensity of NbC film, and $I_z$ is the optical intensity at different sample position $z$ in Z-scan system.

## 3. Normalized optical transmittance of NbC film versus $I_z$

The normalized transmittance versus position sample $z$ under on-focus intensities ($I_0$) of 146, 156, 182, and 208 GW/cm² in Figure 4(a) were also transferred into $I_z$-dependent normalized transmittance. The transferred experimental data can be well fitted by Equation (S7) as shown in Figure S3. Thereafter, the parameters of $\kappa$ and $I_s$ are extracted from fitting results as summarized in Tabel S1.

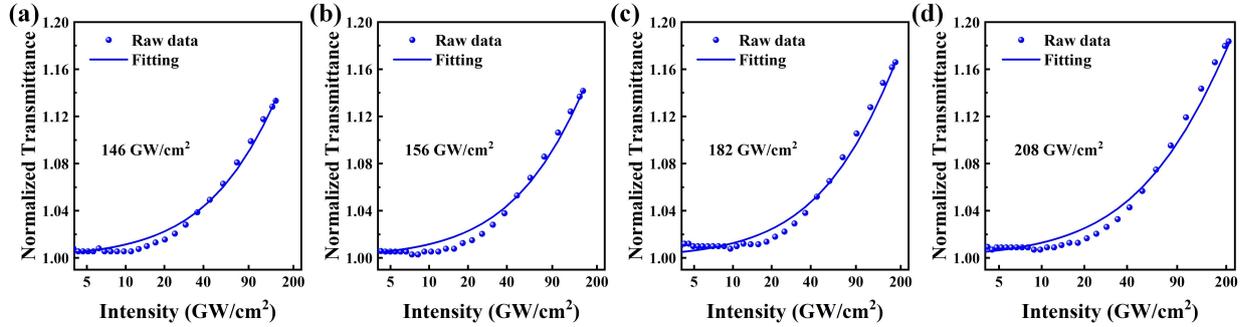

**Figure S3.** Normalized transmittance of NbC film versus $I_z$ under $I_0$ of (a) 146 GW/cm², (b) 156 GW/cm², (c) 182 GW/cm² and (d) 208 GW/cm².

**Table S1.** Optical parameters of NbC film under different $I_0$.

| $I_0$ (GW/cm²) | $\beta_{SA}$ (cm/GW) | Im $\chi^3$ (×10⁻⁸ esu) | FOM (×10⁻¹³ esu cm) | $\kappa$ | $I_s$ (GW/cm²) |
|---|---|---|---|---|---|
| 146 | -350 | -1.53 | 2.27 | 0.72 | 149 |
| 156 | -342 | -1.49 | 2.21 | 0.65 | 152 |
| 182 | -333 | -1.45 | 2.16 | 0.55 | 153 |
| 208 | -312 | -1.36 | 2.03 | 0.54 | 151 |
| 234 | -307 | -1.34 | 1.99 | 0.28 | 154 |

## 4. Relationship between nonlinear absorption coefficient of NbC and $I_0$

For the saturable absorption response, $\alpha_{SA}(I)$ has been described in Equation (S6). Besides, $\alpha_{SA}(I)$ has the general expression:

$$\alpha_{SA}(I) = \alpha_0^{SA} + \beta_{SA} I \tag{S8}$$

Where $\beta_{SA}$ is the nonlinear absorption coefficient of NbC film.

Combining Equation (S6) with (S8), the relationship between $\beta_{SA}$ and $I_0$ is express as:

$$\beta_{SA} = (\kappa - 1)\alpha_0^{SA}/(I_0 + I_s) \tag{S9}$$

Considering the influence of experimental error, the parameters of 'a' and 'b' are introduced. The final relationship between $\beta_{SA}$ and $I_0$ is described as follows:

$$\beta_{SA} = a(\kappa - 1)\alpha_0^{SA}/(bI_0 + I_s) \tag{S10}$$

Where, $\kappa$ can be expressed as $\kappa = 1.35 - 0.0043 \times I_0$ according to $\kappa$ values versus $I_0$ in Table S1.

## 5. Energy-level model for two-photon absorption (TPA) in GaS film

Due to photon energy (1.55 eV) of excitation light lower than the GaS bandgap (2.29 eV), the valence band electrons of GaS need absorb two photons to be excited into conduction band. As such, the TPA energy-level model is established as shown in inset of Figure 5(c). The corresponding rate equations are expressed as follows:

$$\frac{\partial N_0}{\partial t} = -\sigma_{TPA} N_0 \frac{I^2}{2h\nu} + \frac{N_1}{\tau_{10}} \tag{S11}$$

$$\frac{\partial N_1}{\partial t} = \sigma_{TPA} N_0 \frac{I^2}{2h\nu} - \frac{N_1}{\tau_{10}} \tag{S12}$$

$$N_{RSA} = N_0 + N_1 \tag{S13}$$

$$\frac{dI}{dz'} = -\sigma_{TPA} N_0 I^2 - \sigma_{12} N_1 I \tag{S14}$$

$$\frac{dI}{dz'} = -\alpha_{RSA}(I)I \tag{S15}$$

Where $N_0$ ($N_1$) is the carrier population densities in the ground state (excited state) of GaS; $N_{RSA}$ is total carrier population densities of GaS; $\sigma_{TPA}$ is the TPA cross-section; $\sigma_{12}$ is the excited state; $\tau_{10}$ is the relaxation time of electrons from excited state ($S_1$) to ground state ($S_0$); $I$ is the optical intensity and $z'$ is the transmission position of the light in the sample.

Combining the Equation (S11)-(S15), the total nonlinear absorption coefficient of GaS film can be expressed as:

$$\alpha_{RSA}(I) = N_{RSA} \frac{\left(\sigma_{TPA}I + \sigma_{12}\frac{I^2}{I_{s,2pa}^2}\right)}{\left(1 + \frac{I^2}{I_{s,2pa}^2}\right)} \quad (S16)$$

Where $I_{s,2pa}^2 = \frac{2h\nu}{\sigma_{TPA}\tau_{10}}$.

Thus, the normalized transmittance of GaS film can be expressed as:

$$T_{RSA} = [1 - \alpha_{RSA}(I)L_{RSA}] / [1 - \alpha_0^{RSA}L_{RSA}] \quad (S17)$$

Here, $L_{RSA}$ and $\alpha_0^{RSA}$ are thickness and linear absorption coefficient of GaS film.

Substituting the Equation (S16) into (S17), the normalized transmittance expression versus $I_z$ is as follows.

$$T_{RSA}(I_z) = \left(1 - N_{RSA} \frac{\left(\sigma_{TPA}I_z + \sigma_{12}\frac{I_z^2}{I_{s,2pa}^2}\right)}{\left(1 + \frac{I_z^2}{I_{s,2pa}^2}\right)}\right) / \left(1 - \alpha_0^{RSA}L_{RSA}\right) \quad (S18)$$

## 6. Normalized optical transmittance of GaS versus $I_z$

The $I_z$-dependent normalized transmittance of GaS, transferred from Z-scan data under $I_0$ of 250, 280, 312, and 350 GW/cm$^2$ in Figure 5(a) is fitted by Equation (S18) as shown in Figure S4. The fitting results reveal that the light-GaS interaction is dominated by TPA process and exhibit the parameters of $\sigma_{TPA}$ and $\sigma_{12}$ as summarized in Table S2.

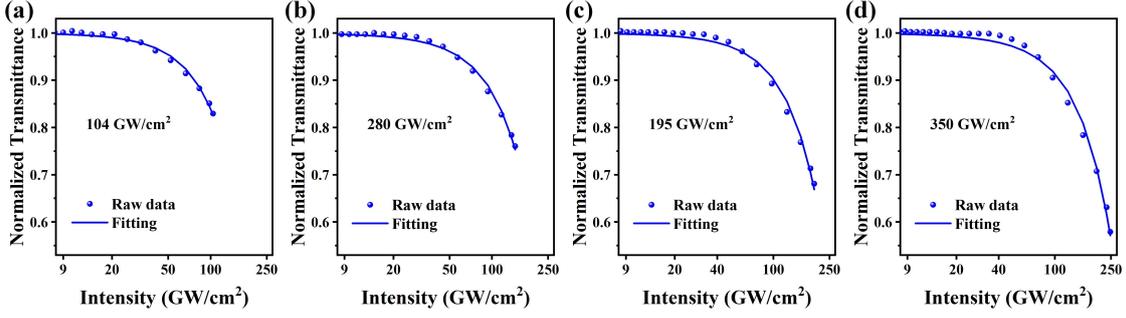

**Figure S4.** Normalized optical transmittance of GaS film versus $I_z$ under $I_0$ of (a) 250 GW/cm$^2$, (b) 280 GW/cm$^2$, (c) 312 GW/cm$^2$, and (d) 350 GW/cm$^2$.

**Table S2.** Optical parameters of GaS film under different $I_0$.

| $I_0$ (GW/cm$^2$) | $\beta_{RSA}$ (cm/GW) | Im $\chi^3$ (esu) | FOM (esu cm) | $\sigma_{TPA}$ (cm$^4$/GW) | $\sigma_{12}$ (cm$^2$) | $\sigma_{TPA}/\sigma_{12}$ (cm$^2$/GW) |
|---|---|---|---|---|---|---|
| 250 | 831 | 1.10×10$^{-7}$ | 1.43×10$^{-12}$ | 0.71×10$^{-27}$ | 5.28×10$^{-17}$ | 0.32×10$^{-10}$ |
| 280 | 812 | 1.07×10$^{-7}$ | 1.40×10$^{-12}$ | 1.86×10$^{-27}$ | 6.68×10$^{-17}$ | 0.77×10$^{-10}$ |
| 312 | 800 | 1.06×10$^{-7}$ | 1.38×10$^{-12}$ | 3.41×10$^{-27}$ | 1.84×10$^{-17}$ | 1.47×10$^{-10}$ |
| 350 | 776 | 1.03×10$^{-7}$ | 1.34×10$^{-12}$ | 5.29×10$^{-27}$ | 3.51×10$^{-17}$ | 1.54×10$^{-10}$ |
| 375 | 752 | 0.99×10$^{-7}$ | 1.29×10$^{-12}$ | 7.65×10$^{-27}$ | 4.07×10$^{-17}$ | 1.88×10$^{-10}$ |

## 7. Relationship between nonlinear absorption coefficient of GaS and $I_0$

When taking no account to the excited state absorption in TPA energy-level model, total nonlinear absorption coefficient can be expressed as:

$$\alpha_{RSA}(I) = \sigma_{TPA} N_{RSA} I \bigg/ \left(1 + \frac{I^2}{I_{s,2pa}^2}\right) \tag{S19}$$

In TPA absorption process, there doesn't include linear absorption. Thus, the general expression of $\alpha_{RSA}(I)$ is as follows.

$$\alpha_{RSA}(I) = \beta_{RSA} I \tag{S20}$$

Combining Equation (S19) with (S20), the relationship between $\beta_{RSA}$ and $I_0$ is express as:

$$\beta_{RSA} = \sigma_{TPA} N_{RSA} \bigg/ \left[1 + \left(I_0/I_{s,2pa}\right)^2\right] \tag{S21}$$

Considering the influence of experimental error, the parameters of 'c' is introduced. The

Equation (S21) is described as:

$$\beta_{RSA} = \sigma_{TPA} N_{RSA} \Big/ \left[ 1 + \left( cI_0 / I_{s,2pa} \right)^2 \right] \tag{S22}$$

Due to the $\sigma_{TPA}$ is dependent on $I_0$, $\sigma_{TPA}$ can be expressed as $\sigma_{TPA} = (0.05 \times I_0 - 13.25) \times 10^{-27}$, according to $\sigma_{TPA}$ values versus $I_0$ in Table S2.

## 8. Numerical simulation of all-optical diode

The nonreciprocal transmission properties of all-optical diode can be simulated by sequentially solving the Equation (1-3) in the main text using the fourth-order Runge-Kutta method. In the forward bias, the pump light first passes through the saturable absorption (SA) material. The Equation (1) in the main text is solved to obtain the intensity output from NbC, which is used as input intensity of interface in NbC/GaS heterostructures. Subsequently, the Equation (3) in the main text is used to describe the light propagation process in the interface of NbC/GaS heterostructures. Finally, Equation (2) in the main text is used to solve the optical intensity output form the reverse saturable absorption (RSA) material. In the reversed bias, the pump light interacts with RSA material, interface, and SA material in turn. Therefore, the order of solving Equation (1-3) in the main text is reversed.

To precisely determine the variation of optical intensity within the nonlinear optical materials under the influence of linear and nonlinear absorption, the NbC/GaS sample were divided into plenty of spatial slices with the interval of 1 nm. whenever the pump light passes through a spatial slice, the Equation (1-3) in the main text were solved by taking advantage of fourth-order Runge-Kutta method to obtain the output intensity which is used as the input intensity for the next spatial slice.

The detailed process of solving differential equations using four-order Runge-Kutta method based on finite-difference scheme is as follows[3]. The generalized differential equation for propagating the optical intensity in the nonlinear two-dimensional materials can be described as:

$$\frac{dI}{dz'} = f(z', I) \tag{S23}$$

The equations describing the fourth-order Runge-Kutta method are expressed as follows.

$$I_{i+1} = I_i + \frac{1}{6}(k_1 + 2k_2 + 2k_3 + k_4)h \tag{S24}$$

$$k_1 = f(z_i', I_i)$$

$$k_2 = f\left(z_i' + \frac{1}{2}h, I_i + \frac{1}{2}k_1 h\right)$$

$$k_3 = f\left(z_i' + \frac{1}{2}h, I_i + \frac{1}{2}k_2 h\right)$$

$$k_4 = f(z_i' + h, I_i + k_3 h)$$

Where $h$ is the step size of $z'$ ($h=1$ nm in simulation); $i$ is the number of spatial slices and $k_m$ ($m=1,2,3,4$) is $m$-order slope over the interval $h$.

## 9. Van der Waals heterostructure energy-level model in NbC/GaS heterostructures

Due to the metallic property of NbC and semiconductor property of GaS, the Schottky heterojunction is formed at the interface between NbC and GaS films. The energy-level of the heterostructures model is shown in Figure 1(b) and the corresponding rate equations are as follows:

$$\frac{dN_0}{dt} = -\sigma_{TPA} N_0 \frac{I^2}{2hv} + \frac{N_1}{\tau_1} + \frac{N_s}{\tau_3} \tag{S25}$$

$$\frac{dN_1}{dt} = \sigma_{TPA} N_0 \frac{I^2}{2hv} - \frac{N_1}{\tau_1} - \frac{N_1}{\tau_2} \tag{S26}$$

$$\frac{dN_e}{dt} = \sigma_s N_s \frac{I}{hv} - \frac{N_e}{\tau_4} + \frac{N_1}{\tau_2} \tag{S27}$$

$$\frac{dN_s}{dt} = -\sigma_s N_s \frac{I}{hv} + \frac{N_e}{\tau_4} - \frac{N_s}{\tau_3} \tag{S28}$$

$$N = N_0 + N_1 + N_s + N_e \tag{S29}$$

$$\frac{dI}{dz'} = -\sigma_{TPA} N_0 I^2 - \sigma_{12} N_1 I - \sigma_s N_s I - \sigma_e N_e I = -\alpha(I)I \tag{S30}$$

Where $\alpha(I)$ is the absorption coefficient of the NbC/GaS heterostructures; $\tau_1$ and $\tau_4$ are the first excited-state lifetimes of GaS and NbC; $\tau_2$ and $\tau_3$ are transition times from first excited-state of GaS to that of NbC and from ground state of NbC to that of GaS and other parameters are same as the parameters mentioned in above model.

Substitute Equation (S25-S29) into Equation S(30), the propagation equation of the interface in NbC/GaS heterostructure can be expressed as:

$$\frac{dI}{dz'} = -N \frac{\left[\sigma_{TPA}I + \sigma_{12}\frac{I^2}{I_1} + \sigma_s\left(\frac{I^2}{I_2} - \frac{I^2}{I_1}\frac{\tau_3}{\tau_1}\right) + \sigma_e\left(\frac{I}{I_3} + \frac{\tau_4}{\tau_3}\right)\left(\frac{I^2}{I_2} - \frac{I^2}{I_1}\frac{\tau_3}{\tau_1}\right)\right]}{\left[1 + \frac{I^2}{I_1} + \left(\frac{I^2}{I_2} - \frac{I^2}{I_1}\frac{\tau_3}{\tau_1}\right) + \left(\frac{I}{I_3} + \frac{\tau_4}{\tau_3}\right)\left(\frac{I^2}{I_2} - \frac{I^2}{I_1}\frac{\tau_3}{\tau_1}\right)\right]} I \quad (S31)$$

Where $I_1 = \frac{2h\nu}{\sigma_{TPA}\tau_1} + \frac{2h\nu}{\sigma_{TPA}\tau_2}$, $I_2 = \frac{2h\nu}{\sigma_{TPA}\tau_3}$, and $I_3 = \frac{h\nu}{\sigma_s\tau_4}$.